\documentclass[twocolumn,showpacs,preprintnumbers,amsmath,amssymb]{revtex4}
\usepackage{graphicx}
\usepackage{dcolumn}
\usepackage{bm}

\begin{document}
\draft
\tighten
\title{Relativistic Photon Mediated Shocks}
\author{Amir Levinson \& Omer Bromberg}
\address{Raymond and Beverly Sackler School of Physics and Astronomy\\ Tel Aviv University,
Tel Aviv 69978, Israel}
\begin{abstract}
A system of equations governing the structure of a steady, relativistic radiation dominated shock
is derived, starting from the general form of the transfer equation obeyed by 
the photon distribution function.  Closure is obtained by truncating the system of moment equations
at some order.  The anisotropy of the photon distribution function inside the shock is shown to 
increase with increasing shock velocity, approaching nearly 
perfect beaming at upstream Lorentz factors $\Gamma_{-}>>1$.  Solutions of the shock equations 
are presented for some range of upstream conditions.  These solutions are shown to converge as
the truncation order is increased.  
\end{abstract}

\maketitle
\narrowtext
Relativistic shocks play an important role in essentially all classes of high-energy
compact astrophysical systems.  Behind these shocks the bulk energy of a collimated
outflow expelled from the central engine is dissipated and converted to the
radio-through-$\gamma$-ray emission observed.   The micro-physics of the shock depends
on the conditions in the upstream and downstream regions.  Shocks that form in a region of small optical depth for Thomson scattering 
are mediated by collective plasma processes, and are termed collisionless shocks \cite{BE87}.  Various observations 
suggest that these shocks may provide sites for acceleration of nonthermal particles (both electrons and protons)
and, at least in some systems (e.g., GRB afterglow), may somehow generate macroscopic, sub-equipartition magnetic fields
in the downstream region \cite{GW99}.  The latter processes may in fact be an inherent part of the shock structure \cite{KKW07}.
The micro-physics of collisionless shocks is highly involved.  Recent efforts have led to important progress in the study of collisionless 
shocks, e.g., Refs. \cite{KKW07,ML99,LE06,Sp07,CSA07}, but despite these efforts many key issues remain as yet unresolved. 
   
When the ratio of radiation to matter pressure in the post shock region exceeds a certain level the shock transition may become 
mediated by photons.  Such conditions are anticipated in, e.g., internal \cite{Eic94,EL00} and oblique \cite{BL07} 
shocks producing the GRB prompt emission, during shock breakout in supernovae and hypernovae \cite{MM99,MW01}, in blazars, and 
in accretion flows onto a black hole \cite{BP81c,Ti97,LT01}.
The effect of radiation on the structure of a non-relativistic, strong shock
has been studied by different authors \cite{ZR67,Weaver76,Becker98,BP81a,BP81b,LS82,Rf88} and applied to various astrophysical systems, such as 
supernovae \cite{Weaver76} and accretion flows onto neutron stars \cite{Becker98}.
The structure and transmitted photon spectrum of a cold, radiation mediated shock was computed in Ref.~\cite{BP81b}, where it was shown
that bulk Comptonization on the converging flow can give rise to a power law spectrum at high energies with a spectral index that tends 
to unity for large Mach numbers.  The model has been generalized later to incorporate thermal effects \cite{LS82,Rf88}.  
The analysis in Ref~\cite{BP81b} is restricted to sufficiently optically thick shocks with small upstream 
velocity, viz.,  $\beta_{-}<<1$.  In this regime the scattered photon distribution is nearly isotropic at every point inside the shock
and the diffusion approximation can be employed to solve the transport equation describing the evolution of the photon population 
across the shock \cite{BP81a}.   Moreover, to order $\beta_{-}^2$ the radiation field
satisfies the equation of state $P_{rad} = U_{rad}/3$, which provides a closure condition for
the set of hydrodynamic equations governing the shock structure.  This simplifies the analysis considerably, 
but renders its applicability to most high-energy compact sources of little relevance. 

Several complications arise in the relativistic regime.  Firstly,
the optical depth across the shock, $\tau \sim c/\beta_{-}$, approaches
unity as the shock becomes relativistic, hence photons do not experience
multiple scattering.  This is a consequence of the fact that 
the average energy a photon gain in a single scattering is large, viz., $\Delta \epsilon/\epsilon >1$.
As a result, the photon distribution function across the shock is anticipated to be
highly anisotropic, rendering the diffusion approximation inapplicable.
Obtaining a closure of the hydrodynamic shock equations then 
becomes an involved problem.  Secondly, the optical depth of a fluid slab 
having a Lorentz factor $\Gamma>1$ depends on the angle $\theta$ between the photon 
direction and the shock velocity as $d\tau=\Gamma(1-\beta\cos\theta)dx$, and needs to be properly 
accounted for.  Thirdly, pair creation may become important if the photon energy exceeds the 
pair creation threshold.  Sufficiently relativistic shocks may therefore
become dominated by electron-positron pairs. From 
the physical point of view, then, a relativistic radiation-dominated shock 
is a complex physical phenomenon, posing an intriguing unsolved problem.  In what follows
we derive a coupled set of hydrodynamic equations that provide a complete description of the shock 
structure, starting from the general form of the transfer equation obeyed by the photon
distribution function.  We then propose a method to close the resultant system of equations and 
present solutions for some range of upstream conditions.  Convergence of these solutions is verified.

The fluid in the shock transition layer is a mixture of baryons, e$^\pm$ pairs, and radiation. 
We denote by $u^\alpha=(\Gamma,\Gamma{\bf \beta})$ the 4-velocity of the mixed fluid, by $n_b$ and $n_\pm$ the number
density of baryons and pairs, respectively (the total number of electrons in this notation is 
$n_e=n_b+n_-$), and by $T^{\mu\alpha}_b$, $T^{\mu\alpha}_\pm$, and  $T^{\mu\alpha}_r$  the 
stress-energy tensors of baryons pairs and radiation, respectively.
The entire system must conserve energy, momentum and baryon number:
\begin{eqnarray}
& &\frac{\partial}{\partial x_\alpha}\left(n_b^\prime u^\alpha\right)=0,\label{eq:cont}\\
& &\frac{\partial}{\partial x_\alpha}\left(T^{\mu\alpha}_b+T^{\mu\alpha}_\pm+T^{\mu\alpha}_r \right)=0.
\label{eq:Tunu}
\end{eqnarray}
The energy-momentum tensor of the radiation can be expressed explicitly in terms of the photon distribution
function, $f_r(k,x^\mu)$, as
\begin {equation}
T^{\mu\nu}_r(x^\mu)=\int{k^\mu k^\nu f_r(k,x^\mu)\frac{d^3k}{k^0}}, 
\end{equation}
where $k^\mu$ denotes the 4-momentum of a photon.  The distribution function satisfies a transfer 
equation, which we express in the form \cite{HS76}
\begin{eqnarray}
&&k^\mu\frac{\partial f_r}{\partial x^\mu}=n^\prime_{l}\int{R(k,k_i)f_r(k_i)\frac{d^3k_i}{k_i^0}}\label{eq:transfer}\\
&&-n^\prime_{l}\int{R(k_i,k)f_r(k)\frac{d^3k_i}{k_i^0}}+C_{pp}(f_r,f_\pm,k)+S_k.\nonumber
\end{eqnarray}  
Here $n^\prime_{l}=n^\prime_b+n^\prime_++n^\prime_-$ is the net number density of scatterers (electrons plus positrons), 
as measured in the fluid rest frame (henceforth, primed quantities refer to the fluid rest frame), $R(k,k_i)$ is the 
redistribution function for Compton scattering from initial state $k_i$ to a final state $k$, the operator $C_{pp}(f_r,f_\pm,k)$ accounts 
for the change in $f_r$ due to e$^\pm$ pair creation and annihilation, and $S_k$ is a source term
associated with all other processes that create or destroy photons 
(e.g., free-free emission and absorption). Conservation of energy, momentum and number of 
quanta of the interacting pair-photon system implies
\begin {eqnarray}
& &\frac{\partial}{\partial x_\alpha}T^{\mu\alpha}_\pm=-\int{k^\mu C_{pp}(f_r,f_\pm,k)\frac{d^3k}{k^0}},\label{eq:T_pm}\\
& &\frac{\partial}{\partial x_\alpha}\left(n^\prime_\pm u^\alpha\right)=-\int{C_{pp}(f_r,f_\pm,k)\frac{d^3k}{k^0}}\label{eq:n_pm}.
\end{eqnarray}
The above set of equations augmented by appropriate boundary conditions upstream provides a complete
description of the shock transition layer.  
To simplify the problem we ignore KN effects in eq. (\ref{eq:transfer}).  We then have
\begin{equation}
R(k^\prime,k^\prime_i)=\frac{3\sigma_T}{16\pi}[1+(\hat{{\bf k}^\prime}\cdot \hat{{\bf k}_i^\prime})^2]
\delta(k^{\prime 0}-k_i^{\prime 0}),\label{eq:redist}
\end{equation}
where $\sigma_T$ is the cross section for Thomson scattering.  Multiplying Eq. (\ref{eq:transfer}) 
by $k^\nu$, integrating over final states $k$, using Eq. (\ref{eq:redist}), and relating $k^{\prime}$
and $k$ through the Lorentz transformation $k^\nu=\Lambda^{\nu}_{\lambda}k^{\prime\lambda}$, we obtain
\begin {equation}
\frac{\partial}{\partial x^\mu}T^{\mu\nu}_b=-n^\prime_l\sigma_T\{\Lambda^{\nu}_{0}T_r^{\prime00}-
\Lambda^{\nu}_{\lambda}T_r^{\prime\lambda 0}\}=n^\prime_l\sigma_T \Lambda^{\nu}_{k}T_r^{\prime k 0}.
\label{eq:motion1}
\end{equation}

We consider a plane-parallel shock in the ($y,z$) plane (the flow is moving in $x$-direction), and 
solve the above equations in the shock frame, where the system is assumed to be in a steady state.
For the range of upstream conditions explored here the downstream temperature is well below $m_ec^2$,
and so the pairs are sub-relativistic.  To a good approximation we then have $T_\pm^{\mu\nu}=n_\pm m_ec^2u^\mu u^\nu$.
Since the shock is radiation dominated we
can neglect the pressure contributed by the particles to get $T_b^{\mu\nu}=m_pc^2n_b^\prime u^\mu u^\nu$. 
Using the latter result and the transformation $T_r^{\prime 0x}=\Lambda^{0}_{\mu}(-\beta)\Lambda^{x}_{\nu}(-\beta)T^{\mu\nu}$,
the zeroth component of Eq. (\ref{eq:motion1}) yields an equation of motion for the fluid Lorentz factor:
\begin{equation}
J\frac{d}{d\tau} \Gamma= \beta T_r^{\prime 0x}=\Gamma^2\beta \left[(1+\beta^2)T_r^{0x}-\beta\left(T_r^{00}+T_r^{xx}\right)\right],
\label{eq:motion2}
\end{equation}
where $d\tau=\sigma_T n^\prime_l\Gamma dx$ is the angle-averaged optical depth for Thomson scattering 
of a fluid slab of thickness $dx$, and $J=m_pc^2n_b^\prime u^x$ is the integral of Eq. (\ref{eq:cont})  .  To 
complete the set of equations a closure relation is needed.
Far upstream ($x=-\infty$) and downstream ($x=+\infty$) the radiation field is in equilibrium, and the
equation of state $3T^{\prime xx}_r=T^{\prime 00}$ applies.  However, within the shock transition layer
this relation no longer holds.   It is convenient to define a dimensionless function $\xi$ that measures the
deviation from isotropy at any given point:
\begin{equation}
T^{\prime xx}_r=\frac{1}{3}\left(1+\xi\right) T_r^{\prime 00}.
\label{eq:xi}
\end{equation}
\begin{figure}[h]
\includegraphics[width=9.0cm]{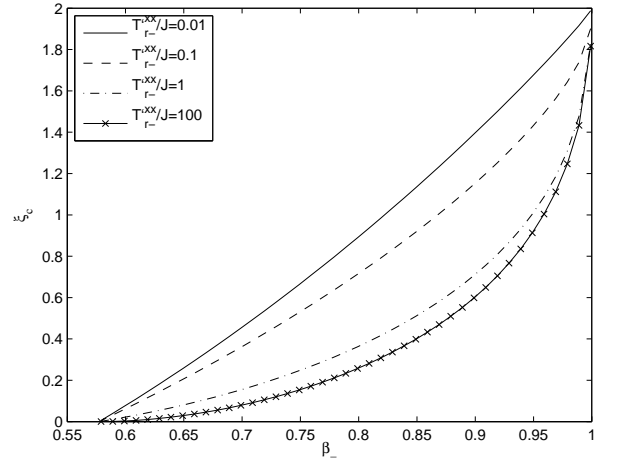}
\caption{$\xi_c$ versus upstream velocity $\beta_{-}$, for different values of the ratio of radiation pressure
and rest mass energy density, as indicated}
\end{figure}
In general $\xi$ is a function of the shock velocity and its derivatives, that is, $\xi=\xi(u^\mu,u^\mu_{,\nu})$.
Owing to relativistic boosting the radiation field is expected to be 
beamed preferentially in the direction of fluid motion.  Consequently we anticipate values between $\xi=0$ 
(complete isotropy) and $\xi=2$ (perfect beaming in the $x$ direction) at any given point in the shock transition layer.
Let us denote by $C_1$ and $C_2$ the integrals of 
the zeroth and $x$ components of Eq. (\ref{eq:Tunu}), respectively.  These integration constants are determined
by the conditions upstream the shock, at $x=-\infty$.  From Eqs. (\ref{eq:cont}), (\ref{eq:Tunu}) and (\ref{eq:xi}) 
we obtain the net photon flux in the local rest frame of the fluid:
\begin{eqnarray}
&&T_r^{\prime 0x}=-\frac{1+\xi}{1+\xi-3\beta^2}[\frac{J}{\Gamma}+m_ec^2n_\pm\beta \label{eq:flux-com}\\
&&-\frac{C_1\left(1+\xi+3\beta^2\right)}{1+\xi}+\frac{C_2\left(1+\xi+3\right)\beta}{1+\xi}].\nonumber
\end{eqnarray}
It can be readily shown that for $\xi=0$ the flux $T_r^{\prime 0x}$ has a root at $\beta=\beta_{+}$, and
$\beta=\beta_{-}$, where $\beta_{+}(\beta_{-})$ is the velocity of the downstream fluid, as determined from the 
shock jump conditions.  It is also seen that the comoving photon flux has a singular point at $\beta^2=(1+\xi)/3$.
Since $T_r^{\prime 0x}$ must be finite everywhere inside the shock it implies that the 
numerator on the R.H.S of Eq. (\ref{eq:flux-com}) must vanish at the critical point.  
For upstream velocity $\beta_{-}>1/\sqrt{3}$ we can then readily constrain $\xi$.  For illustration, we 
present in Fig. 1 plots of $\xi_c$, the value of $\xi$ at the singularity, 
for cases where the energy density of pairs can be neglected in Eq. (\ref{eq:flux-com}). The presence of pairs
is not expected to change the results significantly.  The maximum value of $\xi$ inside the shock must lie between 
$\xi_c$ and $2$.  As seen from Fig. 1, $\xi$ approaches the maximal value $\xi=2$
as the shock becomes relativistic, implying a highly beamed radiation field inside the shock.

In the limit $\beta_{-}<<1$ we expand $\Gamma$ in powers of $\beta$ and $T_r^{\prime 0x}$ 
in powers of $\beta$ and $\xi$.  
By employing Eqs. (\ref{eq:flux-com}) and (\ref{eq:motion2}) and noting that the average photon energy is 
well below the pair production threshold, in which case we can set $n_\pm=0$, 
we recover, to order O($\beta^2$), the result derived in Ref.~\cite{BP81b}:
\begin{equation}
\frac{d}{d\tau} \mu=\frac{7}{2}\mu^2-4(1+\pi_{-})\mu+\frac{1}{2}+4\pi_{-},
\end{equation}
where $\mu=\beta/\beta_{-}$ is the normalized fluid velocity and $\pi_{-}=(T_r^{xx}/J\beta)_{-}$ is the ratio of 
radiation pressure and ram pressure far upstream.   It can also be readily shown that to this order $T^{0x}_r=
dT^{xx}_r/d\tau$, implying that in this limit photon transport is indeed a diffusion process.

In the relativistic case the particle content of the shocked fluid may be dominated by $e^\pm$ pairs.  The pair density 
can be computed from Eq. (\ref{eq:n_pm}) once the collision 
term $C_{pp}$ is known.  However, the latter depends on the energy distribution of photon, the calculation of which is beyond the 
scope of this paper.  As a rough estimate of the pair density we use the equilibrium value,
\begin{equation}
n^\prime_\pm=8\pi(2\pi)^{1/2}\left(\frac{m_ec}{h}\right)^3\left(\frac{kT}{m_ec}\right)^{3/2}\exp(-m_ec^2/kT).
\label{eq:n_pm_eqi}
\end{equation}
To complete our treatment we need to derive an equation describing the change in $\xi$ across the shock.  We do so by
integrating Eq. (\ref{eq:transfer}) over $k^0$ and $\phi$.  
Defining $I(\mu)=\int{(k^{0})^3fdk^0d\phi}$, and likewise $I^\prime(\mu^\prime)=\int{(k^{\prime 0})^3f^\prime dk^{\prime 0}d\phi^\prime}$,
recalling that $I(\mu)=\Gamma^4(1+\beta\mu^\prime)^4 I^\prime(\mu^\prime)$, and noting that $d\mu^\prime/d\tau=(\mu^\prime-1)
(\Gamma\beta)^{-1}d\Gamma/d\tau$, we arrive at,
\begin{eqnarray}
&&\Gamma^2(\beta+\mu^\prime)\frac{d I^\prime}{d \tau}+4\Gamma\frac{d\Gamma}{d\tau}\mu^\prime(1+\mu^\prime/\beta)I^\prime
+I^\prime\label{eq:transfer2}\\ 
&&=\frac{T^{\prime00}}{2}\left[1+\frac{(3\mu^{\prime 2}-1)}{8}\xi\right]+\kappa_{pp}(\mu^\prime),\nonumber
\end{eqnarray}
\begin{figure}[h]
\includegraphics[width=9.0cm]{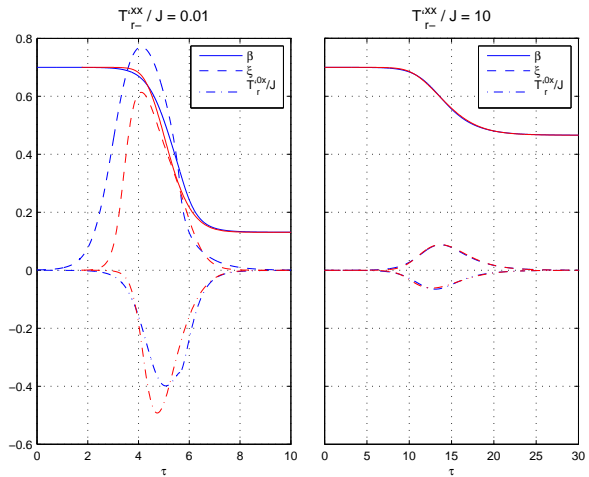}
\caption{Fluid velocity, comoving photon flux $T_{r}^{\prime 0x}$, and anisotropy parameter $\xi$ as functions of
optical depth, for matter dominated (left panel) and radiation dominated (right panel) fluid far upstream.  The blue 
lines in both panels are solutions obtained to second order and the red lines to third order (see text for further details)}
\end{figure}
where $n^\prime_l\sigma_T\kappa_{PP}=\Gamma^3(1-\beta\mu)^3\int{(k^{0})^2 C_{PP} dk^0d\phi}$.  
We solve Eq. (\ref{eq:transfer2}) by expanding $I^\prime$ and $\kappa_{PP}$ in terms of Lagendre polynomials:
\begin{equation}
I^\prime(\mu^\prime)=\frac{1}{2}\Sigma{\eta_kP_k(\mu^\prime)};\ \ \ \kappa_{pp}(\mu^\prime)=
\frac{1}{2}\Sigma{\zeta_kP_k(\mu^\prime)}.
\label{eq:poly}
\end{equation}
\begin{figure}[h]
\includegraphics[width=9.0cm]{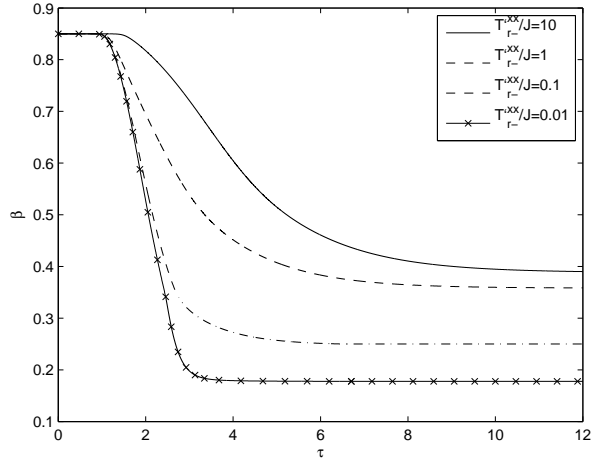}
\caption{Velocity profile for an upstream fluid with $\Gamma_{-}=2$ and 
different values of $T_{r}^{\prime xx}/J$.}
\end{figure}
Note that with this normalization $T_r^{\prime 00}=\eta_0$, $T_r^{\prime 0x}=\eta_1/3$ and $\xi=2\eta_2/5\eta_0$.  
Substituting Eq. (\ref{eq:poly})
into Eq. (\ref{eq:transfer2}) yields an infinite set of ODE's for the unknown variables $\eta_n$.
To simplify the analysis we keep only the terms $\zeta_0$ and $\zeta_1$ in the expansion for $\kappa_{pp}$
which, using Eq. (\ref{eq:T_pm}), can be expressed as $\zeta_0=\Gamma^2(\beta dT^{xx}_\pm/d\tau-dT^{0x}_\pm/d\tau)$
and $\zeta_1=\Gamma^2(\beta dT^{0x}_\pm/d\tau-dT^{xx}_\pm/d\tau)$\footnote{For the solutions of mildly relativistic shocks 
presented below we find that the energy density of pairs is anyhow negligible}.  A closure relation can be 
obtained by truncating the expansion of $I^\prime(\mu^\prime)$ in Eq. (\ref{eq:poly}) at some order $n$.   It can be 
shown that the resultant set of equations has singularities at certain values of $\beta$, that are independent of the
conditions upstream.   The root in the denominator of Eq. (\ref{eq:flux-com}), which can alternatively be derived
using the moment equations, is one of these points.   As seen, the full transfer equation, 
Eq. (\ref{eq:transfer2}), has a singularity only at the point 
$\beta=-\mu^\prime$ (corresponding to $\mu=0$ in the shock frame), where the optical depth tends to infinity.  This implies that
any physical shock solution of Eq. (\ref{eq:transfer2}) must automatically pass through the singular points of the moment 
equations.  This is demonstrated already in Eq. (\ref{eq:flux-com}) where the solution for $\xi$ in 
all cases studied automatically takes the value at the singularity that leaves 
$T_r^{\prime 0x}$ finite.   These singular points are therefore different in some respects from the critical points of a hydrodynamic
system that fix physical parameters of a transonic solution (see Ref.~\cite{Nobili91} for a detailed account).
We were able to find physical shock solutions of the moment equations that pass smoothly through these singular points, for mildly 
relativistic upstream velocities ($\Gamma_{-}\sim$ a few).
Full details will be given elsewhere.  For the cases studied below we find that the solutions converge as 
the order of truncation is increased.  An example is shown in Fig. 2, where the fluid velocity $\beta$, 
comoving photon flux $T_{r}^{\prime 0x}$, and anisotropy parameter $\xi$ are plotted as functions of the angle
averaged optical depth, $\tau$, for upstream velocity $\beta_{-}=0.7$.  Second order (blue lines) and third order (red lines)
solutions are compared in each panel.  As seen, for the radiation dominated upstream condition (right panel)
the agreement is perfect.  For the matter dominated condition the convergence is slower.  This is in part a consequence of 
the larger anisotropy (larger values of $\xi$ and $T^{\prime0x}_r$) in the latter example.  We have obtained also
higher order solutions for this case and found nearly complete convergence at fourth order.   Shock profiles computed
for upstream Lorentz factor $\Gamma_-=2$ are exhibited in fig 3.
The ultra-relativistic regime requires analysis of higher order terms and is currently under study.

From the above analysis it is evident that an appreciable fraction of the shock energy should be converted, via 
bulk Comptonization, to high energy photons having energies well in excess of the thermal peak.  
If the optical depth in the upstream region is not too large then these photons would escape before being thermalized.
Consequently, {\em a non-thermal spectral component appears to be an inherent feature of relativistic radiation dominated shocks and 
requires no particle acceleration, as in the case of collisionless shocks}.
The transmitted spectrum should extend up to an energy of $\sim m_ec^2$, as measured in the 
shock frame, above which it will be suppressed by KN effects.  In the frame of the observer the high energy cutoff of the 
emitted spectrum will be boosted by a factor $\Gamma_s$, the shock Lorentz factor.
For GRBs this implies cutoff energy that can exceed 50 MeV or so in the observer frame.
This mechanism can therefore account for the spectra observed in most GRBs.
Detailed calculations of the transmitted spectrum are underway (Bromberg \& Levinson, in preparation).  

Relativistic photon mediated shocks can provide a means for producing nonthermal spectra also in purely leptonic fireballs, 
the consideration of which is motivated by recent post-SWIFT discoveries of a shallow afterglow phase at early times
\cite{Burrows05}.  According to some interpretations (e.g., Ref.~\cite{Fan05})
these observations indicate prolonged activity of the central engine that, if true, 
implies that $\gamma$-rays are emitted during the prompt phase with very high efficiency. 
Such episodes can be most naturally explained as resulting from photon mediated shocks in a pure 
electron-positron plasma.  Further discussions can be found in Refs.~\cite{BL07,Le06}

As seen from Figs 2 the scale of a photon mediated shock is a few Thomson depths, which for parameters
typical to most compact, relativistic systems is several orders of magnitudes larger than any microphysical 
scale (e.g., the skin depth and Larmor radius) associated with
collisionless shocks.  Thus, unlike collisionless shocks, photon mediated shocks cannot accelerate particles
to nonthermal energies.  This has direct implications for production of cosmic rays and VHE neutrinos in
those systems.  A particular example is the recent proposal that failed GRBs may be prodigious sources of 
TeV neutrinos \cite{Razz04}.  The idea is based on the {\em ad hoc} assumption that
internal collisionless shocks that form in the chocked outflow accelerate protons to very high energies.  However, 
the large Thomson depth anticipated in the region where the shocks form \cite{Razz04} should render them radiation dominated,
as our analysis indicates, and so they may not be able to provide the required sites for the acceleration of protons.
Thus, the effects of radiation domination are likely to have a strong
influence on the predictions presented in \cite{Razz04}.

We thank Yuri Lyubarsky and the anonymous referee for useful comments.  This research was 
supported by  an ISF grant for a Israeli Center for High Energy Astrophysics.


\begin{thebibliography}{99}
\bibitem{BE87} R.~D. Blandford \& D. Eichler, Phys. Rep. {\bf 154}, 1 (1987)
\bibitem{GW99} A. Gruzinov \& E. Waxman,  Astrophys. J. {\bf 511}, 852 (1999)
\bibitem{KKW07} B. Katz, U. Keshet, \& E. Waxman, Astrophys. J. {\bf 655}, 375 (2007)
\bibitem{ML99} A.~V.~Medvedev \& A. Loeb, Astrophys. J. {\bf 526}, 697 (1999)
\bibitem{LE06} Y. Lyubarski \& D. Eichler,  Astrophys. J. {\bf 647}, 1250 (2006)
\bibitem{Sp07} A. Spitkovsky Astrophys. J. Lett., submitted (2007; arXiv:0706.3126)
\bibitem{CSA07} P. Cheng, A. Spitkovsky \& J. Arons, Astrophys. J., in press (2007; arXiv:0704.3832)
\bibitem{Eic94} D. Eichler, Astrophys. J. Supp., {\bf 90}, 877 (1994)
\bibitem{EL00} D. Eichler \& A. Levinson,  Astrophys. J. {\bf 529}, 146 (2000)
\bibitem{BL07} O. Bromberg \& A. Levinson,  Astrophys. J., {\bf 671}, 678 (2007)
\bibitem{MM99} C.~D. Metzner \& C.~F. McKee,  Astrophys. J. {\bf 510}, 379 (1999)
\bibitem{MW01} P. Meszaros \& E. Waxman, Phys. Rev. Lett. {\bf 87}, 171102 (2001)
\bibitem{BP81c} R.D.~Blandford, \& D.G.~Payne, Man. Not. R. Astron. Soc., {\bf 196}, 781 (1981c)
\bibitem{Ti97} L.~G. Titarchuk, Astrophys. J. {\bf 487}, 834 (1997)
\bibitem{LT01} P. Laurent \& L.~G. Titarchuk, Astrophys. J. Lett., {\bf 562}, L67 (2001)
\bibitem{ZR67} Zel'dovich, Ya. B. \& Raizer, Yu. P. 1967, Physics of Shock Waves and 
   High-Temperature Hydrodynamic Phenomena, Academic Press, New York
\bibitem{Weaver76} T.A.~Weaver,  Astrophys. J. Supp., {\bf 32}, 233 (1976)
\bibitem{Becker98} P.A.~Bekker,  Astrophys. J., {\bf 327}, 772 (1988)
\bibitem{BP81a} R.D.~Blandford, \& D.G.~Payne, Man. Not. R. Astron. Soc., {\bf 194}, 1033 (1981a)
\bibitem{BP81b} R.D.~Blandford, \& D.G.~Payne, Man. Not. R. Astron. Soc., {\bf 194}, 1041 (1981b)
\bibitem{LS82} Y.E. Lyubarsky, \& R.~A. Sunyaev, Soviet Astron. Lett, {\bf 8}, 330 (1982)
\bibitem{Rf88} Riffert, H., ApJ, 327, 760 (1988)
\bibitem{HS76} S.~H. Hsieh \& E.~A. Spiegel, Astrophys. J. {\bf 207}, 244 (1976)
\bibitem{Nobili91} L. Nobili, R. Turolla, \& L. Zampieri, Astrophys. J. {\bf 383}, 2250 (1991) 
\bibitem{Burrows05} D.~N. Burrows, et al. {\it Science}, {\bf 309}, 1833 (2005);
S. Barthelmy, et al. {\it Nature}, {\bf 438}, 994 (2005)
\bibitem{Fan05} Y.~Z. Fan, and D.~M. Wei {\it Mon. Not. R. Astron. Soc.}, {\bf 364}, L42 (2005)
\bibitem{Le06} A. Levinson, Int. J. Mod. Phys. A, {\bf 21}, 6015 (2006)
\bibitem{Razz04} S. Razzaque, P. Meszaros, \& E. Waxman, PRL, {\bf 93}, 181101 (2004); 
S. Razzaque, P. Meszaros, \& E. Waxman, Mod. Phys. Lett. A {\bf 20}, 2351 (2005)
\end{thebibliography}
\end{document}